\documentclass[a4paper,12pt]{iopart}

\usepackage{rotating}

\usepackage{graphicx}
\usepackage{dcolumn}
\usepackage{bm}

\begin{document}

\title[Direct measurement of the  {\rm $^{15}$N(p,$\gamma$)$^{16}$O} total cross section]{Direct measurement of the  $^{15}$N(p,$\gamma$)$^{16}$O total cross section at novae energies}

\author{D Bemmerer$^1$\footnote{e-mail: \mailto{d.bemmerer@fzd.de}}, A Caciolli$^{2,3}$, 
R~Bonetti$^4$\footnote{deceased}, C~Broggini$^2$, F~Confortola$^5$, P~Corvisiero$^5$, H~Costantini$^5$, Z~Elekes$^6$, A~Formicola$^7$, Zs~F\"ul\"op$^6$, G~Gervino$^{8}$, A~Guglielmetti$^4$,
C~Gustavino$^7$, Gy~Gy\"urky$^6$, 
M~Junker$^7$, B~Limata$^9$, M~Marta$^1$, 
R~Menegazzo$^2$, P~Prati$^5$, V~Roca$^9$, C~Rolfs$^{10}$, C~Rossi~Alvarez$^2$, E~Somorjai$^6$, and O~Straniero$^{11}$ \\
(The LUNA collaboration)
}

\address{$^1$ Forschungszentrum Dresden-Rossendorf, Dresden, Germany}
\address{$^2$ Istituto Nazionale di Fisica Nucleare (INFN), Sezione di Padova, Italy}
\address{$^3$ Universit\`a di Padova, Padova, Italy}
\address{$^4$ Istituto di Fisica Generale Applicata, Universit\`a di Milano and INFN Sezione di Milano, Italy}
\address{$^5$ Universit\`a di Genova and INFN Sezione di Genova, Genova, Italy}
\address{$^6$ Institute of Nuclear Research (ATOMKI), Debrecen, Hungary}
\address{$^7$ INFN, Laboratori Nazionali del Gran Sasso (LNGS), Assergi (AQ), Italy}
\address{$^{8}$ Dipartimento di Fisica Sperimentale, Universit\`a di Torino and INFN Sezione di Torino, Torino, Italy}
\address{$^9$ Dipartimento di Scienze Fisiche, Universit\`a di Napoli "Federico II" and INFN Sezione di Napoli, Napoli, Italy}
\address{$^{10}$ Institut f$\ddot{\mathrm{u}}$r Experimentalphysik III, Ruhr-Universit$\ddot{\mathrm{a}}$t Bochum, Germany}
\address{$^{11}$ Osservatorio Astronomico di Collurania, Teramo, and INFN Sezione di Napoli, Napoli, Italy}

\date{\today}

\begin{abstract}
The $^{15}$N(p,$\gamma$)$^{16}$O reaction controls the passage of nucleosynthetic material from the first to the second carbon-nitrogen-oxygen (CNO) cycle. A direct measurement of the total $^{15}$N(p,$\gamma$)$^{16}$O cross section at energies corresponding to hydrogen burning in novae is presented here. Data have been taken at 90 -- 230\,keV center-of-mass energy using a windowless gas target filled with nitrogen of natural isotopic composition and a bismuth germanate summing detector. 
The cross section is found to be a factor two lower than previously believed.
\end{abstract}


\section{Introduction}

\begin{figure}[tb]
\centering
 \includegraphics[angle=00,width=0.5\columnwidth]{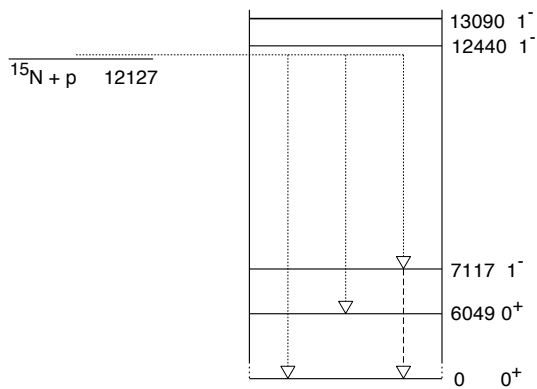}
 \caption{\label{fig:Levels-O16} Energy levels of $^{16}$O relevant to the $^{15}$N(p,$\gamma$)$^{16}$O reaction at low energy, in keV \cite{Tilley_16-17}. Primary (dotted) and secondary $\gamma$-ray transitions are also shown.}
\end{figure}

The $^{15}$N(p,$\gamma$)$^{16}$O reaction ($Q$-value $Q$ = 12.127\,MeV) links the CN cycle \cite{Bethe39-PR_letter,Weizsaecker38-PZ} to the CNO bi-cycle and all further CNO cycles \cite{Iliadis07}. 
The $^{15}$N(p,$\gamma$)$^{16}$O cross section $\sigma(E)$ ($E$ denotes the center of mass energy in keV, $E_{\rm p}$ the proton beam energy in the laboratory system) can be parameterized \cite{Iliadis07} by the astrophysical S-factor $S(E)$ defined as
\begin{equation} \label{eq:Sfactor}
S(E) = \sigma(E) E \exp(212.85/\sqrt{E}).
\end{equation}
At astrophysically relevant energies $E$ $<$ 1\,MeV, the $^{15}$N(p,$\gamma$)$^{16}$O excitation function is influenced by two resonances at $E_{\rm p}$ = 335 and 1028\,keV ($E_{\rm x}$ = 12440 and 13090\,keV, figure~\ref{fig:Levels-O16}), with respective widths of  $\Gamma_{\rm p}$ = 91 and 130\,keV, both decaying predominantly into the ground state of $^{16}$O \cite{Tilley_16-17}. For the $E_{\rm x}$ = 12440 (13090) keV level, 1.2\% (0.58\%)  decay branching to the 0$^+$ first excited state of $^{16}$O at 6.049\,MeV has been reported \cite{Tilley_16-17}. In addition, for the 13090\,keV level, there is 3.1\% decay branching to the 1$^-$ third excited state at 7.117\,MeV \cite{Tilley_16-17}. No other decays to $^{16}$O excited states are known for $E_{\rm p}$ $\leq$ 1028\,keV \cite{Tilley_16-17}. 

The non-resonant cross section has been studied in previous experiments using NaI \cite{Hebbard60-NP} and Ge(Li) \cite{Rolfs74-NPA} detectors, reporting cross section data for  $E_{\rm p}$ = 150 -- 2500\,keV  \cite{Rolfs74-NPA}. Citing discordant normalizations between those two studies \cite{Hebbard60-NP,Rolfs74-NPA}, only the data from one of these studies \cite{Rolfs74-NPA} have been used in reaction rate compilations \cite{CF88-ADNDT,NACRE99-NPA}. 

Recently, the asymptotic normalization coefficients (ANC's) for direct capture to the ground and several excited states in $^{16}$O have been measured \cite{Mukh08-PRC}. It was found that the low-energy non-resonant yield is dominated by ground state capture \cite{Mukh08-PRC}, but the new ANC leads to a much lower  direct capture cross section (sum of direct capture to all states in $^{16}$O) than previously \cite{Rolfs74-NPA}. The new ANC values have then been used in an R-matrix fit \cite{Mukh08-PRC} including also the cross section data from refs.~\cite{Hebbard60-NP,Rolfs74-NPA}, suggesting a factor two lower astrophysical S-factor than previously believed \cite{Rolfs74-NPA,CF88-ADNDT,NACRE99-NPA}. Another recent R-matrix analysis concentrating on ground state capture was based again on the direct data from refs.~\cite{Hebbard60-NP,Rolfs74-NPA}, and it also indicates a much lower S-factor \cite{Barker08-PRC}. In view of the conflicting data \cite{Hebbard60-NP,Rolfs74-NPA} and the recent extrapolations \cite{Mukh08-PRC,Barker08-PRC}, new experimental data is clearly called for.

The aim of the present work is to experimentally determine the $^{15}$N(p,$\gamma$)$^{16}$O cross section directly at energies corresponding to hydrogen burning in novae. The relevant temperatures in novae \cite{Jose98-ApJ,Jose07-ApJL} are $T_6$ = 200 - 400 ($T_6$ denoting the central temperature of a star in units of 10$^6$ K), corresponding to Gamow energies \cite{Iliadis07} of $E_{\rm Gamow}$ = 150-240\,keV. In order to obtain the new cross section data, spectra from a radiative proton capture experiment at LUNA that has been performed using nitrogen gas of natural isotopic composition (99.6\% $^{14}$N, 0.4\% $^{15}$N) have now been analyzed regarding the $^{15}$N(p,$\gamma$)$^{16}$O reaction. 

\section{Experiment}
\label{sec:Experimental}

The experiment has been performed at the Laboratory for Underground Nuclear Astrophysics (LUNA) in Italy's Gran Sasso underground laboratory (LNGS). The LUNA facility has been designed for measuring low nuclear cross sections for astrophysical purposes \cite{Bonetti99-PRL,Casella02-NPA,Formicola04-PLB,Lemut06-PLB,Bemmerer06-PRL,Marta08-PRC,Formicola08-NPA3}, benefiting from its ultra-low laboratory $\gamma$-ray background \cite{Bemmerer05-EPJA,Caciolli08-arxiv}. 

\subsection{Target}

A windowless, differentially pumped gas target cell filled with 1\,mbar nitrogen gas of natural isotopic composition (0.366\% $^{15}$N \cite{Coplen02-PAC}) has been irradiated with $E_{\rm p}$ = 100 -- 250\,keV H$^+$ beam from the 400\,kV LUNA2 accelerator \cite{Formicola03-NIMA}.  The emitted $\gamma$-rays have been detected in a 4$\pi$ BGO summing crystal \cite{Casella02-NIMA}. The calorimetric beam intensity values are known with 1.0\% precision \cite{Casella02-NIMA}. 

The natural isotopic composition of the target gas enabled parallel experiments on $^{14}$N(p,$\gamma$)$^{15}$O \cite{Lemut06-PLB,Bemmerer06-NPA} and $^{15}$N(p,$\gamma$)$^{16}$O (present work). The $^{14}$N(p,$\gamma$)$^{15}$O analysis is already published including full experimental details \cite{Lemut06-PLB,Bemmerer06-NPA}; the present work concentrates on aspects pertinent to obtaining the $^{15}$N(p,$\gamma$)$^{16}$O  cross section. 

During the experiment, nitrogen gas of natural isotopic composition and 99.9995\% chemical purity was flowing through the windowless target cell with a flux of 2 liters/second. No recirculation was used, so the gas was discarded after one passage through the target. The effective $^{15}$N target density for the present work has been obtained scaling the known target density (3.2\% uncertainty including the beam heating correction \cite{Bemmerer06-NPA}) with the standard isotopic composition \cite{Coplen02-PAC}. A recent survey has found that $>$99\% of nitrogen-bearing materials have isotopic abundances within 2.0\% of the standard value \cite{Coplen02-PAC}, which is defined to be that of atmospheric air. The $^{15}$N content of atmospheric air on different continents has been found to be constant to 2.6\% \cite{Mariotti83-Nature}, and commercial tank gas even falls within 1.0\% of the standard \cite{Junk58-GCA}. In order to verify whether these findings also apply to the presently used tank gas, gas samples of the type of nitrogen used here and from the same supplier have been sent to three different laboratories for isotopic analysis. The isotopic ratio was found to be within 3\% of the standard. As relative uncertainty for the isotopic ratio, 3\% is therefore adopted.

\subsection{$\gamma$-ray detection}
\label{subsec:Efficiency}

The $\gamma$-ray detection efficiency of the BGO detector \cite{Casella02-NIMA} has been obtained by a dedicated simulation with GEANT4 \cite{Agostinelli03-NIMA}. The simulation has been validated at low $\gamma$-ray energy by measurements with calibrated $\gamma$-ray sources and at $E_\gamma$ $\approx$ 7\,MeV by a detailed comparison with the results from the previous \cite{Bemmerer06-NPA} GEANT3 simulation. An uncertainty of 3.0\% is quoted here for the probability of detecting isotropically emitted 12\,MeV $\gamma$-rays.  

The GEANT4 summing detector efficiency depends, however, also on inputs from experiment, such as the decay scheme and the angular distribution of the emitted $\gamma$-radiation. If the capture does not proceed directly to the ground state, but to some excited state, several $\gamma$-rays may be emitted, leading to lower detection efficiency when compared to ground state capture. 

In order to understand the decay scheme, germanium spectra taken at $E_{\rm p}$ = 400\,keV (slightly above the  $E_{\rm p}$ = 335\,keV resonance) bombarding solid Ti$^{\rm nat}$N targets with proton beam \cite{Marta08-PRC} have been reanalyzed here. Experimental upper limits of 1.9\% (1.8\%) for the primary $\gamma$-rays for capture to the excited states at 7.117 (6.049) MeV in $^{16}$O have been derived. In addition, from a reanalysis of germanium spectra \cite{Bemmerer05-EPJA} taken with the present gas target setup at  $E_{\rm p}$ = 200\,keV, an upper limit of 6\% for the $\gamma$-ray from the decay of the 7.117\,MeV state is deduced. These findings are consistent with the previous conclusion that for $E_{\rm p}$ $<$ 400\,keV, the reaction proceeds to $\geq$95\% by capture to the ground state in $^{16}$O \cite{Rolfs74-NPA}. 

The GEANT4 simulation shows that the summing peak detection efficiency for $\gamma$-rays decaying through the 1$^-$ level at 7.117\,MeV is 27\% lower than for ground state capture. The 0$^+$ level at 6.049\,MeV does not decay by $\gamma$-emission, so capture to this level cannot be detected in the 12\,MeV summing peak at all. Scaling these effects with the above mentioned experimental upper limits for the capture probability to the corresponding level, 1.9\% systematic uncertainty for the total cross section is obtained due to possible capture to excited states. 

The angular distribution has previously been found to be isotropic at the $E_{\rm p}$ = 1028\,keV resonance \cite{Rolfs74-NPA}, and for the present analysis, isotropy has been assumed. The simulation shows that due to the large solid angle covered by the BGO, the detection efficiency is enhanced by only 4\% when assuming a complete $\sin^2$$\vartheta$ shape instead. In order to account for this effect, 4\% is adopted as systematic uncertainty. 

\begin{figure}[tb]
\centering
 \includegraphics[angle=0,width=\columnwidth]{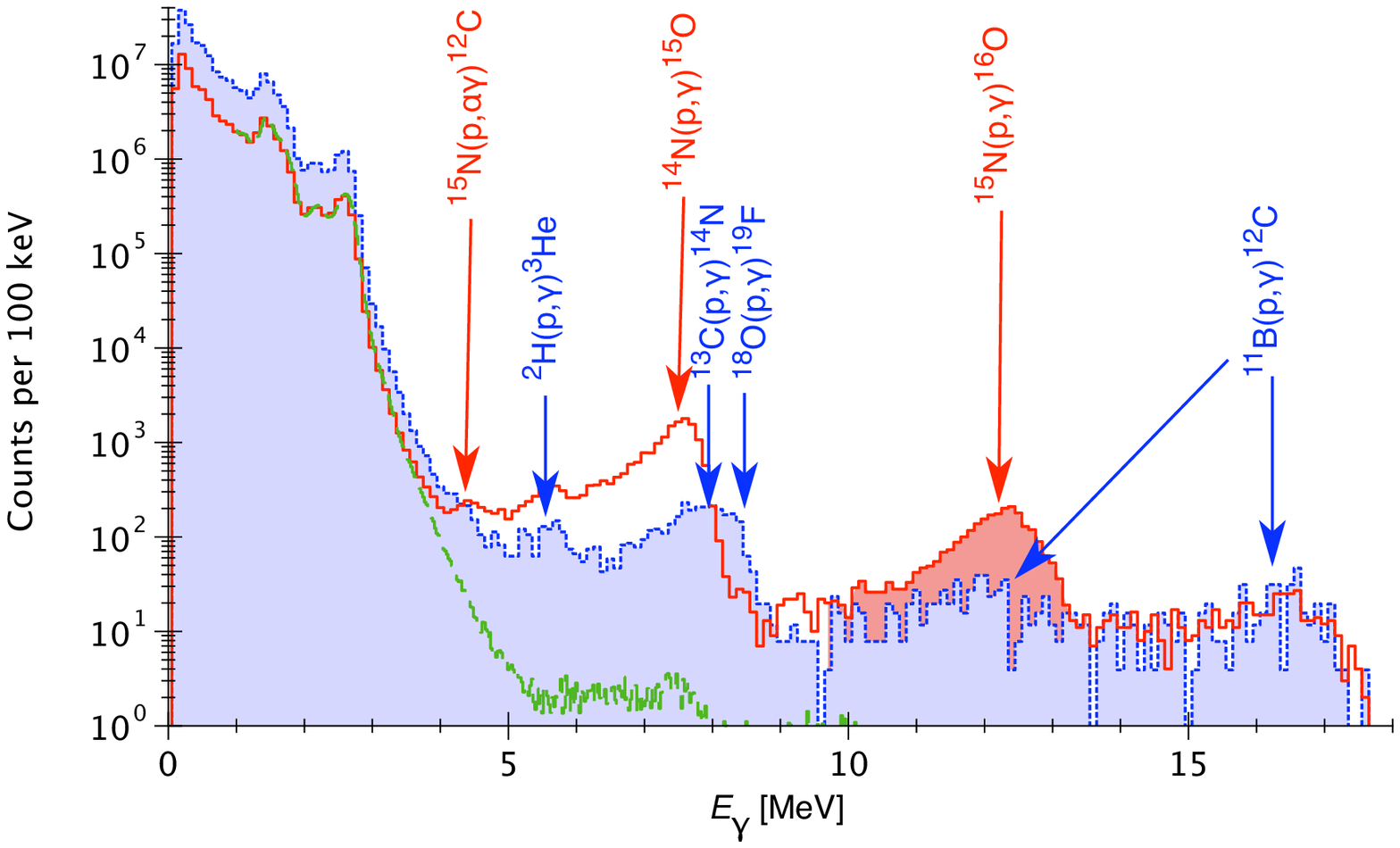}
 \caption{\label{fig:Spec150} $\gamma$-ray spectrum recorded at $E_{\rm p}$ = 150\,keV. Solid red (dotted blue) line: Nitrogen gas in the target (helium gas, rescaled to match the nitrogen spectrum in the 14.4 -- 18.0\,MeV region). Dashed green line, laboratory background, rescaled for equal livetime. See text for details.}
\end{figure}

\subsection{Analysis of the $\gamma$-ray spectra}
\label{subsec:Analysis}

During the experiment, $\gamma$-ray spectra were taken at twelve different incident energies between $E_{\rm p}$ = 100-250\,keV. For each beam energy, two in-beam spectra were recorded: one with 1\,mbar nitrogen gas (natural isotopic composition) in the target, and one with 1\,mbar helium gas (chemical purity 99.9999\%) to monitor ion beam induced background. In addition, a spectrum with 1\,mbar argon gas in the target has been recorded at $E_{\rm p}$ = 216\,keV. Laboratory background spectra were taken during accelerator downtimes. 

\begin{figure}[tb]
\centering
 \includegraphics[angle=90,width=1.0\columnwidth]{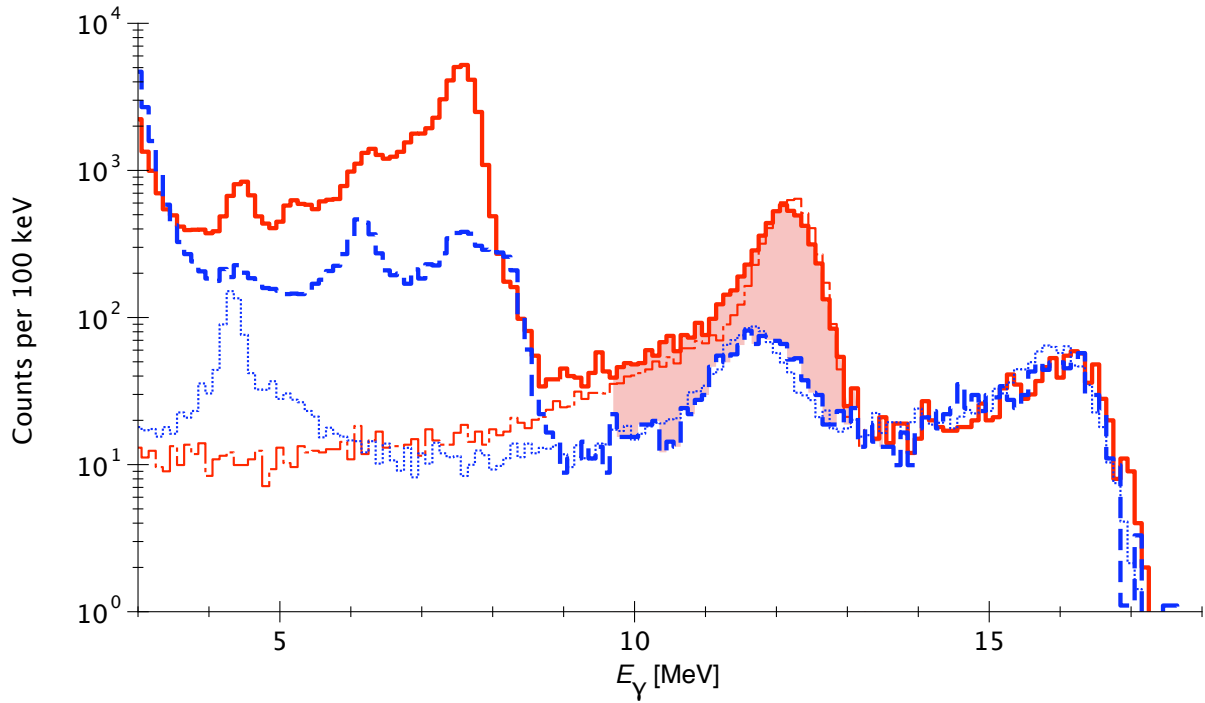}
 \caption{\label{fig:Spec220} $\gamma$-ray  spectrum, $E_{\rm p}$ = 220\,keV. Red solid (dot-dashed) line: Experimental, nitrogen gas (simulated, assuming only the $^{15}$N(p,$\gamma$)$^{16}$O reaction). Blue dashed (dotted) line: Experimental, helium gas, rescaled as in figure \ref{fig:Spec150} (simulated with Sim.\,B, assuming only the $^{11}$B(p,$\gamma$)$^{12}$C reaction).}
\end{figure}

The in-beam spectra can be classified in two groups, low beam energies $E_{\rm p}$ = 100-150\,keV (example, figure \ref{fig:Spec150}), and high beam energies $E_{\rm p}$ = 190-250\,keV (example, figure \ref{fig:Spec220}). Salient features of the spectra are discussed in the following.

At low $\gamma$-ray energies ($E_\gamma$ $\leq$ 4\,MeV), the in-beam $\gamma$-ray spectra are dominated by the laboratory background and resultant pile-up. For 4\,MeV $<$ $E_\gamma$ $\leq$ 8.5\,MeV, the following in-beam $\gamma$-lines are evident \cite{Bemmerer05-EPJA}:
\begin{itemize}
\item the 4.4\,MeV $\gamma$-ray from the decay of the first excited state of $^{12}$C populated both in the $^{11}$B(p,$\gamma$)$^{12}$C and in the $^{15}$N(p,$\alpha$$\gamma$)$^{12}$C reactions (well visible in all the nitrogen spectra, visible in some of the helium spectra),
\item the $\sim$5.5\,MeV peak from the $^{2}$H(p,$\gamma$)$^{3}$He reaction (visible only for $E_{\rm p}$ $\leq$ 150\,keV in both the nitrogen and helium spectra),
\item the 6.1\,MeV $\gamma$-ray from the decay of the second excited state of $^{16}$O populated in the $^{19}$F(p,$\alpha$$\gamma$)$^{16}$O reaction (visible only for $E_{\rm p}$ $\geq$ 180\,keV in the helium spectra),
\item the 6.2\,MeV and 6.8\,MeV secondary $\gamma$-rays and the $\sim$7.5\,MeV summing peak from the $^{14}$N(p,$\gamma$)$^{15}$O reaction (well visible in all the nitrogen spectra, not visible in the helium spectra),
\item the $\sim$7.7\,MeV peak from the $^{13}$C(p,$\gamma$)$^{14}$N reaction (well visible in the helium spectra, covered by the $^{14}$N(p,$\gamma$)$^{15}$O lines in the nitrogen spectra), and
\item the 8.1\,MeV summing peak from the $^{18}$O(p,$\gamma$)$^{19}$F reaction (visible only in a few helium spectra).
\end{itemize}

At $E_\gamma$ $>$ 8.5\,MeV, the laboratory background \cite{Bemmerer05-EPJA} is negligible for the purposes of the present study (figure~\ref{fig:Spec150}). At these high $\gamma$-ray energies, the spectrum  is determined by only two reactions: 
\begin{enumerate}
\item First, the full energy peak of the $^{15}$N(p,$\gamma$)$^{16}$O reaction to be studied, visible in the nitrogen spectra at $E_\gamma$ = $Q$ + $E$ $\approx$ 12.3\,MeV. Because of the rather smeared out response function of the BGO detector to high-energy monoenergetic $\gamma$-rays, a region of interest (ROI) from 9.7-13.5\,MeV (shaded in figs. \ref{fig:Spec150}, \ref{fig:Spec220}) has been adopted. The probability that a 12\,MeV $\gamma$-ray emitted isotropically at the center of the detector leads to a count in this ROI is found to be 77\% in the simulation. 
\item Second, two peaks from the $^{11}$B(p,$\gamma$)$^{12}$C beam-induced background reaction ($Q$ = 15.957\,MeV), visible in both the nitrogen and helium spectra: a summing peak at $E_\gamma$ = $Q$ + $E$ $\approx$ 16\,MeV and the primary ($E_\gamma$ $\approx$ 12\,MeV) $\gamma$-ray from capture to the 4.439\,MeV first excited state in $^{12}$C. (The decay of that state has been discussed above.)
\end{enumerate}
 
\begin{sidewaystable}[h]
\caption{\label{tab:SpectrumIntegration} Spectrum integration and background subtraction. The raw counts in the ROI (9.7-13.5\,MeV) and in the background monitoring region (14.4-18.0\,MeV) are given. For the ratio $R_{12/16}^{\rm Boron}$, the experimental data are from runs with helium gas in the target. The simulations A and B and the adopted uncertainty are explained in the text. The boron background in the ROI (column 8) is obtained by multiplying columns 3 and 7. The net counts in the peak (column 9) are obtained by subtracting column 8 from column 2.}~\\

\begin{tabular}{crrcrrcrrrr}
  & \multicolumn{2}{l}{Raw counts}  & \multicolumn{4}{l}{$R_{12/16}^{\rm Boron}$}  & \multicolumn{1}{l}{Boron background} & \multicolumn{3}{l}{Net counts}\\ \ns\ns\ns
$E_{\rm p}$ & \crule{2} & \crule{4} & \crule{1} & \crule{3} \\ \ns
\ [keV] & 9.7-13.5 & 14.4-18.0 & Experiment & Sim.\,A & Sim.\,B & adopted & \multicolumn{1}{l}{9.7-13.5} & \multicolumn{1}{l}{9.7-13.5} & $\Delta_{\rm stat}$ & $\Delta_{\rm Boron}$  \\ 
\mr
101 & 293 & 69 & 1.9$\pm$0.6 & 0.70 & 1.13 & 1.9$\pm$1.0 & 130$\pm$70 & 164 & 10\% & 43\% \\
122 & 355 & 31 & 3.3$\pm$1.0 & 0.75 & 1.37 & 3.3$\pm$1.0 & 100$\pm$40 & 252 & 7\% & 14\% \\
131 & 662 & 79 & 1.8$\pm$0.5 & 0.77 & 1.33 & 1.8$\pm$1.0 & 140$\pm$80 & 522 & 5\% & 15\% \\
141 & 1703 & 172 & 1.1$\pm$0.2 & 0.80 & 1.44 & 1.1$\pm$1.0 & 190$\pm$170 & 1510 & 3\% & 11\% \\
151 & 2739 & 433 & 1.6$\pm$0.2 & 0.80 & 1.46 & 1.6$\pm$1.0 & 700$\pm$400 & 2047 & 3\% & 21\% \\
188 & 12126 & 2895 & 1.4$\pm$0.3 & 0.81 & 1.46 & 1.4$\pm$1.0 & 4100$\pm$2900 & 8064 & 1\% & 36\% \\
201 & 1300 & 222 & 1.40$\pm$0.05 & 0.84 & 1.52 & 1.4$\pm$1.0 & 310$\pm$220 & 990 & 4\% & 22\% \\
210 & 32569 & 3836 & 1.42$\pm$0.03 & 0.80 & 1.49 & 1.4$\pm$1.0 & 5400$\pm$3800 & 27120 & 1\% & 14\% \\
216 & & & 1.45$\pm$0.09 \footnote[1]{At $E_{\rm p}$ = 216\,keV, argon gas has been used instead of helium.} & 0.80 & 1.45 & & &  &  \\
221 & 6360 & 902 & 1.72$\pm$0.07 & 0.78 & 1.46 & 1.7$\pm$1.0 & 1600$\pm$900 & 4805 & 2\% & 19\% \\
229 & 1930 & 98 & 2.9$\pm$0.2 & 0.80 & 1.43 & 2.9$\pm$1.0 & 280$\pm$100 & 1649 & 3\% & 6\% \\
238 & 1517 & 33 & 2.2$\pm$0.8 & 0.77 & 1.51 & 2.2$\pm$1.0 & 70$\pm$40 & 1443 & 3\% & 2\% \\
250 & 958 & 18 & 6$\pm$4 & 0.81 & 1.38 & 6$\pm$4 & 110$\pm$70 & 847 & 4\% & 8\% \\
\mr
\end{tabular}
\end{sidewaystable}

\subsection{Subtraction of the {\rm $^{11}$B(p,$\gamma$)$^{12}$C} background}
\label{subsec:Boron}

In order to obtain the $^{15}$N(p,$\gamma$)$^{16}$O cross section, the background in the 9.7-13.5\,MeV ROI induced by the $^{11}$B(p,$\gamma$)$^{12}$C reaction must be reliably determined and subtracted. 

The $^{11}$B counting rate can be monitored by the yield in the 14.4-18.0\,MeV region, where no other beam-induced lines are present. This rate varied strongly from run to run, also at the same beam energy, so it was necessary to derive a background subtraction procedure based on data in the same experimental spectrum used also for the cross section determination. Assuming that the place of origin of the $^{11}$B $\gamma$-rays is the collimator at the entrance of the target cell \cite{Bemmerer05-EPJA}, which is hit by the beam halo (0.5-5\% of the beam current on target), the mentioned variation of the $^{11}$B counting rate can be explained with differences in the details of the proton beam focusing from run to run. 

However, even for different absolute $^{11}$B counting rates, the ratio between the $\approx$12\,MeV and $\approx$16\,MeV  $^{11}$B-induced counting rates depends only on the beam energy (due to energy-dependent branching ratios and angular distributions) and not on the focusing. This leads to the definition of the ratio $R_{12/16}^{\rm Boron}$:
\begin{equation} \label{eq:R1216}
R_{12/16}^{\rm Boron} \stackrel{!}{=} \frac{\rm Counts\,(9.7-13.5\,MeV)}{\rm Counts\,(14.4-18.0\,MeV)} \qquad.
\end{equation}

At each beam energy, the quantity $R_{12/16}^{\rm Boron}$ has been determined experimentally from a monitor run with helium gas in the target (table~\ref{tab:SpectrumIntegration}). As a check on the reliability of using helium as monitor gas, at $E_{\rm p}$ = 216\,keV, $R_{12/16}^{\rm Boron}$ has been determined with argon gas instead of helium, with consistent results (table~\ref{tab:SpectrumIntegration}).

The experimental $R_{12/16}^{\rm Boron}$ values are then compared with the results of two GEANT4 simulations called Sim.\,A and Sim.\,B. In both Sim.\,A and Sim.\,B, the known branching ratios and angular distribution of the $^{11}$B(p,$\gamma$)$^{12}$C reaction from ref.~\cite{Cecil92-NPA} are included.

\begin{itemize}
\setlength{\itemindent}{7mm}
\item[{\bf Sim.\,A}] The point of origin of the $^{11}$B $\gamma$-rays  was assumed not to be the final collimator, but the beamstop (table~\ref{tab:SpectrumIntegration}).
\item[{\bf Sim.\,B}] The point of origin of the $^{11}$B $\gamma$-rays  was assumed to be the final collimator as discussed above (table~\ref{tab:SpectrumIntegration}, figure~\ref{fig:Spec220}).
\end{itemize}

For all data points, Sim.\,B is closer to the experimental data than Sim.\,A. However, at the lowest and highest proton beam energies the experimental $R_{12/16}^{\rm Boron}$ values tend to be even higher than the simulated ones from Sim.\,B (table~\ref{tab:SpectrumIntegration}). In order to understand this phenomenon, it should be noted that the simulation results depend strongly on the assumed branching ratios, angular distributions, and angular correlations. The branching ratio is known experimentally also for off-resonant energies \cite{Cecil92-NPA}. However, the angular distribution is only known at the $E_{\rm p}$ = 163\,keV resonance \cite{Cecil92-NPA}. It seems plausible that given this limited input data, the simulation does a better job close to $E_{\rm p}$ = 163\,keV than far away, at the lowest and highest proton beam energies. 

For the actual data analysis, the experimental  $R_{12/16}^{\rm Boron}$ values have been used. In order to err on the side of caution and quote a conservative uncertainty on the adopted $R_{12/16}^{\rm Boron}$ value, for $\Delta R_{12/16}^{\rm Boron}$ either the statistical uncertainty or $\pm$1.0 (an upper limit on the full difference between Sim.\,A and Sim.\,B) was used, whichever is greater (table~\ref{tab:SpectrumIntegration}).


Finally, the $^{11}$B background to be subtracted in the 9.7-13.5\,MeV ROI of the nitrogen spectrum is then obtained by multiplying the counts in the 14.4-18.0\,MeV monitoring region in the same nitrogen spectrum with the experimental $R_{12/16}^{\rm Boron}$ value from the corresponding helium run (table~\ref{tab:SpectrumIntegration}). The uncertainty due to the boron background subtraction has 1.8-43\% effect on the S-factor data, and it dominates the uncertainty for most data points. Two types of runs have been excluded from the present analysis: Runs that show more $^{11}$B background than $^{15}$N yield in the ROI, and runs for which no helium monitor run has been performed. 

\subsection{Further experimental details}

The effective interaction energy has been calculated assuming a constant astrophysical S-factor \cite{Iliadis07} over the typically 10\,keV thick target, leading to 0.7-3.2\% systematic uncertainty including also the accelerator energy calibration \cite{Formicola03-NIMA} uncertainty. All systematic uncertainties are summarized in table~\ref{tab:Uncertainties}.
 
 \begin{table}[tb]
\caption{\label{tab:Uncertainties} Systematic uncertainties and their effect on the S-factor data.}
\setlength{\extrarowheight}{0.1cm}
\begin{tabular}{llc}
\multicolumn{1}{l}{Source of the uncertainty} & Details found in & Effect on S-factor\\
\hline
Target density & Ref.~\cite{Bemmerer06-NPA} & 3.2\%\\
$^{15}$N isotopic ratio & Refs.~\cite{Mariotti83-Nature,Coplen02-PAC} & 3.0\%\\
Beam intensity & Refs.~\cite{Casella02-NIMA,Bemmerer06-NPA} & 1.0\%\\
Effective energy & Ref.~\cite{Formicola03-NIMA} & 0.7\% -- 3.2\%\\
$\gamma$-ray detection efficiency & Sec. \ref{subsec:Efficiency} & 3.0\%\\
$\gamma$-ray capture to excited states & Sec. \ref{subsec:Efficiency} & 1.9\%\\
$\gamma$-ray angular distribution & Sec. \ref{subsec:Efficiency} & 4.0\%\\
$^{11}$B(p,$\gamma$)$^{12}$C background & Sec. \ref{subsec:Boron} & 1.8\% -- 43\%\\
\hline
Total systematic uncertainty: & & 8\% -- 44\%\\
\end{tabular}
\end{table}

\begin{table}[b]
\caption{\label{tab:Results} Effective center-of-mass interaction energy $E^{\rm eff}$, S-factor data, and
relative uncertainties. The systematic uncertainty due to the boron background subtraction has been derived in table~\ref{tab:SpectrumIntegration} and is repeated here (column 5). The boron uncertainty is already included in the total systematic uncertainty given below (column 4).}
\begin{tabular}{rrrrr}
$E^{\rm eff}$ & $S(E^{\rm eff})$ & \multicolumn{3}{c}{$\Delta$$S$/$S$} \\ \cline{3-5}
\ [keV] & [keV barn] & statistical & total systematic & systematic (boron) \\
\hline
90.0 & 38.4 & 14\% & 44\% & 43\%\\ 
109.3 & 44.4 & 11\% & 16\% & 14\% \\ 
118.5 & 47.0 & 6\% & 17\% & 15\% \\ 
127.9 & 55.4 & 3\% & 13\% & 11\% \\ 
136.6 & 57.6 & 4\% & 22\% & 21\% \\ 
173.0 & 72.2 & 2\% & 37\% & 36\% \\ 
183.2 & 86.1 & 4\% & 24\% & 22\% \\ 
192.3 & 83.8 & 1\% & 16\% & 14\% \\ 
202.8 & 85.9 & 2\% & 20\% & 19\% \\ 
210.3 & 99.9 & 3\% & 9\% & 6\% \\ 
219.4 & 110.4 & 3\% & 7\% & 2\% \\ 
230.0 & 120.9 & 5\% & 11\% & 8\% \\
\end{tabular}
\end{table}

\begin{figure}[tb]
\centering
 \includegraphics[angle=0,width=1.0\columnwidth]{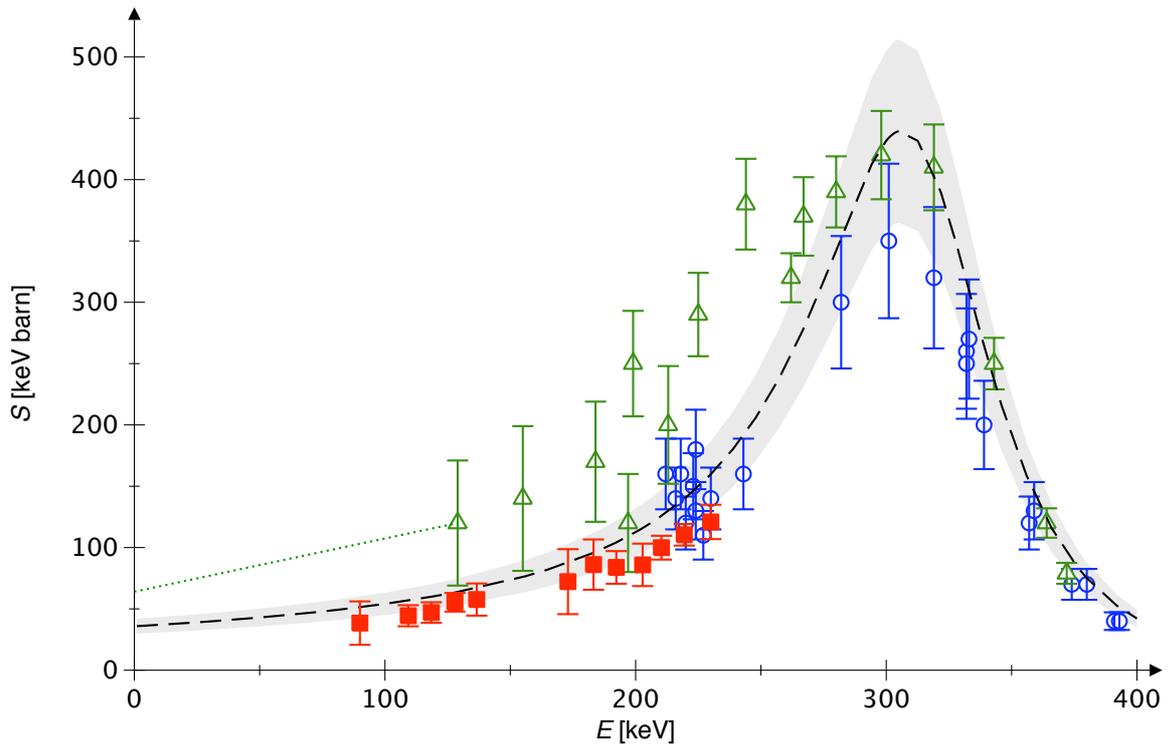}
 \caption{\label{fig:Sfactor} $^{15}$N(p,$\gamma$)$^{16}$O astrophysical S-factor. Experimental data from ref.~\cite{Hebbard60-NP} (blue circles, limited to $E$ $\geq$ 210\,keV), ref.~\cite{Rolfs74-NPA} (green triangles) and the present work (red filled squares). Error bars reflect statistical and systematic uncertainties summed in quadrature. Dotted line, previous low-energy extrapolation by the NACRE compilation \cite{NACRE99-NPA}. Dashed line, previous R-matrix fit, and shaded area, its quoted 17\% uncertainty \cite{Mukh08-PRC}.}
\end{figure}

\section{Results}

Based on the spectrum integration discussed in the previous section, the $^{15}$N(p,$\gamma$)$^{16}$O cross section has been determined at twelve effective center-of-mass interaction energies $E^{\rm eff}$ between 90 and 230\,keV (table~\ref{tab:Results}). The statistical uncertainty is typically well below 10\%. 

The present S-factor data (figure~\ref{fig:Sfactor}) are about a factor two lower than the previous data by ref.~\cite{Rolfs74-NPA}, but still consistent at 2$\sigma$ level given the previous high uncertainties. In the limited overlapping energy region, the present data seem to agree with ref.~\cite{Hebbard60-NP}, if ref.~\cite{Hebbard60-NP}'s low-energy data points (affected by beam-induced background) are excluded. The data from the present work extend to energies lower than ever measured before and are significantly lower than the low-energy extrapolation adopted in the NACRE \cite{NACRE99-NPA} compilation. 

The present data are on average 20\% lower than, but given the previous uncertainty still consistent with, the recent R-matrix fit based on an ANC measurement \cite{Mukh08-PRC}. They are also lower than the fits shown in ref.~\cite{Barker08-PRC}. These R-matrix fits \cite{Mukh08-PRC,Barker08-PRC} had relied on direct experimental data from Refs.~\cite{Hebbard60-NP,Rolfs74-NPA} for the dominating resonant contribution, and it seems prudent to call for a new R-matrix fit, which is beyond the scope of the present work.

Previous one-zone nucleosynthesis calculations of novae \cite{Iliadis02-ApJSS} have shown that a factor two lower $^{15}$N(p,$\gamma$)$^{16}$O rate results in up to 22\% reduction in the final $^{16}$O yield, depending on the nova temperature. Further implications of the changed $^{15}$N(p,$\gamma$)$^{16}$O rate are yet to be studied.

\section{Summary}

The $^{15}$N(p,$\gamma$)$^{16}$O cross section has been measured at energies corresponding to hydrogen burning in novae. The present data are more precise than previous direct experiments \cite{Hebbard60-NP,Rolfs74-NPA}. They are about a factor two lower than the values adopted in reaction rate compilations \cite{CF88-ADNDT,NACRE99-NPA}.

\section*{Acknowledgments}
We thank W. Brand (Max-Planck-Institute for Biogeochemistry Jena, Germany) for assistance with the isotopic abundance analysis.
Financial support by INFN and in part by the European Union (TARI RII3-CT-2004-506222) and the Hungarian Scientific Research Fund (T49245 and K68801)  is gratefully acknowledged.

\section*{References}
\bibliographystyle{iopart-num}
\providecommand{\newblock}{}

\end{document}